\documentclass[preprint,5p,twocolumn]{elsarticle}

\usepackage{amssymb}
\usepackage{graphicx}
\usepackage{xcolor}

\usepackage[nodots]{numcompress}

\def\bsym#1{{\bf #1}}
\def\CRSS{{\rm CRSS}}
\def\MRSSP{{\rm MRSSP}}
\def\diag{{\rm diag}}
\def\gdir#1{\langle #1 \rangle}
\def\gplane#1{\{#1\}}
\def\refeq#1{(\ref{#1})}
\def\reffig#1{Fig.~\ref{#1}}

\journal{Acta Materialia}

\begin{document}

\begin{frontmatter}
  \title{Which stresses affect the glide of screw dislocations in bcc metals?}

  \author{R. Gr\"oger}
  \ead{groger@ipm.cz}
  
  \address{Central European Institute of Technology - Institute of Physics of Materials
    (CEITEC-IPM), \\ Academy of Sciences of the Czech Republic, \v{Z}i\v{z}kova 22, 61662 Brno, Czech
    Republic}
  
  \begin{abstract}
    By direct application of stress in molecular statics calculations we identify the stress
    components that affect the glide of $1/2\gdir{111}$ screw dislocations in bcc tungsten. These
    results prove that the hydrostatic stress and the normal stress parallel to the dislocation line
    do not play any role in the dislocation glide. Therefore, the Peierls stress of the dislocation
    cannot depend directly on the remaining two normal stresses that are perpendicular to the
    dislocation, as proposed by Koester A, Ma A, Hartmaier A. Acta Mater 2012;60:3894 but, instead,
    on their combination that causes an equibiaxial tension-compression (and thus shear) in the
    plane perpendicular to the dislocation line. The Peierls stress of $1/2\gdir{111}$ screw
    dislocations then depends only on the orientation of the plane in which the shear stress
    parallel to the Burgers vector is applied and on the magnitude and orientation of the shear
    stress perpendicular to the slip direction. 
  \end{abstract}
  
  \begin{keyword}
    Peierls stress \sep Screw dislocation \sep Bcc metal \sep Non-glide
    stress \sep Yield criterion.
  \end{keyword}
\end{frontmatter}

%\linenumbers

\section{Introduction}
\label{sec:intro}

It is generally accepted that the plastic deformation of body-centered cubic (bcc) metals is very
different from that of close-packed crystals \cite{christian:83}. The main reason is that the former
is governed by the glide of $1/2\gdir{111}$ screw dislocations whose cores are non-planar
\cite{vitek:70, duesbery:89}. Consequently, the glide of these dislocations may be affected by all
components of the applied stress tensor \cite{ito:01}. These can be divided into two groups. The
first are the stress components that contribute to the Peach-Koehler force on the dislocation and
thus directly cause its glide. The remaining stress components do not exert any Peach-Koehler force
on the dislocation, but they affect the slip by modifying the structure of the dislocation core.

Atomistic simulations made in Ref.~\cite{ito:01} have shown that the Peierls stress of the
$1/2\gdir{111}$ screw dislocation in bcc metals depends on the orientation of the corresponding
maximum resolved shear stress plane (MRSSP) in the zone of its slip direction and on the shear
stress perpendicular to the slip direction. The former relates to the twinning-antitwinning
asymmetry that has been observed in virtually all bcc metals \cite{christian:83,groger:09b}. The
latter represents a non-glide stress that modifies the structure of the dislocation core and thus
increases or decreases the Peierls stress.

We have shown previously using a series of uniaxial loadings \cite{groger:08a} that no other stress
components affect the dislocation glide. This conclusion disagrees with the recent proposal of
Koester et al. \cite{koester:12} that the Peierls stress of bcc iron depends explicitly on three
normal components, two of which are perpendicular and one parallel to the dislocation line. The
objective of this paper is to resolve this controversy by carrying out a series of molecular statics
simulations of an isolated $1/2[111]$ screw dislocation under stress and to identify uniquely the
stress components that affect its glide. Although all these calculations have been made using the
Bond Order Potential (BOP) \cite{horsfield:96, aoki:07} parameterized for bcc tungsten by Mrovec et
al. \cite{mrovec:07}, the conclusions drawn from these studies should be valid generally for all bcc
metals. This is supported by recent calculations made by Chen et al. \cite{chen:13,chen:13a} using a
magnetic BOP for bcc iron \cite{mrovec:11} whose results are very similar to those obtained by us
previously for bcc molybdenum and tungsten \cite{groger:08a}.

\section{Atomistic simulations}

All atomistic simulations carried out in this paper utilized a tetragonal simulation block whose
orientation, and thus the coordinate system in which the loading was applied, were defined
by the Cartesian axes $x=[\bar{1}2\bar{1}]$, $y=[\bar{1}01]$, and $z=[111]$. Periodic boundary
conditions were imposed along the $z$ direction to simulate a straight infinite $1/2[111]$ screw
dislocation, while the widths of the block in the $x$ and $y$ directions were about $30a$, where
$a=3.1652$~\AA{} is the lattice parameter of bcc tungsten. The dislocation was inserted into the
middle of the block by shifting all atoms according to the anisotropic linear-elastic strain field
of the dislocation that was derived originally by Eshelby, Read and Shockley \cite{eshelby:53a}
(for details and further references, see Chapters 13-3 or 13-7 in Hirth and Lothe
\cite{hirth:82}). The atoms in the region $10 < |x|/a \leq 15$ and $10 < |y|/a \leq 15$, called
hereafter the inactive part, were then held fixed while the atoms in the region for which $|x|/a \leq
10$ and $|y|/a \leq 10$ (active part) were relaxed by molecular statics using the BOP for bcc
tungsten \cite{mrovec:07}.

The most general stress tensor that can be applied to the block contains six independent
components. Two of these are the shear stresses parallel to the slip direction -- one
($\sigma_{13}$) acts in the $(\bar{1}2\bar{1})$ plane, while the other ($\sigma_{23}$) in the
$(\bar{1}01)$ plane. The remaining four stress components are the shear stress perpendicular to the
slip direction ($\sigma_{12}$) and the three normal stress components in the direction of the
coordinate axes or, equivalently, the hydrostatic stress $\sigma_h$, the normal stress parallel to
the dislocation line ($\sigma_{33}$) and one of the remaining two stress components ($\sigma_{11}$
or $\sigma_{22}$). 

In the following, we will consider that the MRSSP coincides with the $(\bar{1}01)$ plane. This means
that the stress component $\sigma_{13}$ is zero and the value of $\sigma_{23}$ at which the
dislocation starts to move is the critical resolved shear stress (CRSS). Our objective is to
investigate the dependence of the CRSS on the hydrostatic stress $\sigma_h$, the normal stress
$\sigma_{33}$ parallel to the dislocation line, and the collective effect of the remaining two
stress components, $\sigma_{11}$ and $\sigma_{22}$. For completeness, we also study the effect
  of $\sigma_{12}$ and explain how it affects the CRSS when the MRSSP deviates away from the
  $(\bar{1}01)$ plane.

\subsection{Dependence of the CRSS on individual stresses}

\begin{figure}[!htb]
  \centering
  \includegraphics[scale=0.33]{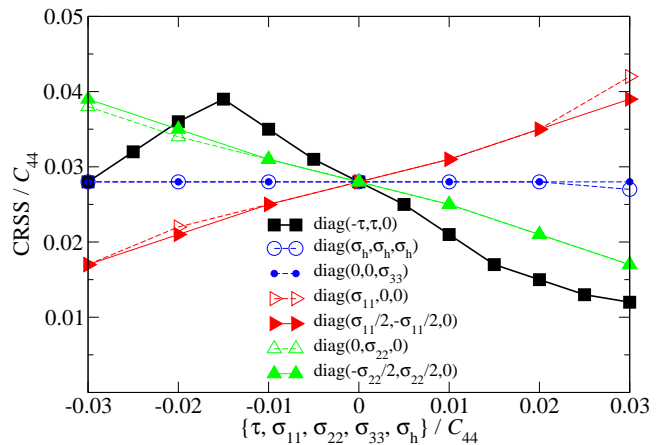}
  \caption{Dependence of the CRSS on individual components of the stress tensor. The black symbols
    represent the CRSS vs. $\tau$ dependence published previously in Ref.~\cite{groger:08a}; this is
    further refined here to allow for a point-wise comparison of the CRSS for an
    equibiaxial tension-compression in the $xy$ plane with the CRSS vs. $\sigma_{11}$ and the CRSS
    vs. $\sigma_{22}$ dependencies.}
  \label{fig:CRSS-all}
\end{figure}

We begin by investigating the dependence of the CRSS on the hydrostatic stress. Starting with the
relaxed atomic block with the dislocation, we further displace all atoms by the stress tensor
$\bsym{\Sigma}_h = \diag(\sigma_h,\sigma_h,\sigma_h)$, where the arguments of $\diag$ are the three
components along the principal diagonal of $\bsym{\Sigma}_h$. We obtained relaxed atomic blocks
for the values $\sigma_h=\{0, \pm0,01, \pm0.02, \pm0.03\}C_{44}$ in which the atoms in the outer
(inactive) region of the block remain displaced by the superposition of the displacement fields of
the dislocation and the stress tensor $\bsym{\Sigma}_h$. In order to investigate the dependence of
the CRSS on the hydrostatic stress, we have subsequently superimposed on each of these relaxed
blocks the stress tensor
\begin{equation}
  \bsym{\Sigma}_{23} = \left[
    \begin{array}{ccc}
      0 & 0 & 0 \\
      0 & 0 & \sigma_{23} \\
      0 & \sigma_{23} & 0
    \end{array} 
    \right] \ .
  \label{eq:Sigma23}
\end{equation}
The stress $\sigma_{23}$ was increased from zero in steps of $0.001C_{44}$, while keeping the stress
$\sigma_h$ fixed. The value of $\sigma_{23}$ at which the dislocation starts to move is then
identified with the CRSS. The obtained CRSS vs. $\sigma_h$ data, plotted in \reffig{fig:CRSS-all}
by blue circles, show that the CRSS is independent of the hydrostatic stress.

We now perform an analogous calculation to investigate the dependence of the CRSS on the stress
component $\sigma_{33}$, i.e. the normal stress that acts parallel to the dislocation line. Starting
with the relaxed atomic block with the dislocation, we apply the stress tensor
$\bsym{\Sigma}_{33}=\diag(0,0,\sigma_{33})$ for the same six values of $\sigma_{33}$ used when
studying the role of the hydrostatic stress $\sigma_h$. The dependence of the CRSS on the stress
component $\sigma_{33}$ is then obtained by superimposing on a fixed stress tensor
$\bsym{\Sigma}_{33}$ the stress tensor \refeq{eq:Sigma23} in steps until $\sigma_{23}$ reaches the
CRSS. The obtained CRSS vs. $\sigma_{33}$ data, plotted in \reffig{fig:CRSS-all} by blue
dots, shows that the CRSS does not depend on the stress component $\sigma_{33}$.

Based on their recent molecular statics calculations on bcc iron, Koester et al. \cite{koester:12}
argued that the CRSS depends on all three normal components along the principal diagonal of the
stress tensor. It was shown in Refs.~\cite{groger:09b, chen:13, chen:13a} that the $1/2\gdir{111}$
screw dislocations in different bcc metals respond similarly to the applied load and thus the
conclusions made in Ref.~\cite{koester:12} should be applicable also to bcc tungsten. However, this
is not the case because we do not observe any dependence of the CRSS on $\sigma_h$ and
$\sigma_{33}$. This implies that the only diagonal stress tensor that may affect the CRSS is
$\bsym{\Sigma}_\tau = \diag(-\tau,\tau,0)$. This stress tensor imposes a shear stress perpendicular
to the slip direction, which can be seen more clearly after rotating the coordinate system by
$-45^\circ$ in the zone of the $z$ axis (i.e. in the direction from the $(\bar{1}01)$ plane towards
the $(0\bar{1}1)$ plane). The dependence of the CRSS on $\tau$, published already in
Ref.~\cite{groger:08a}, is plotted in \reffig{fig:CRSS-all} by black filled squares.

To further demonstrate that the CRSS does not depend independently on the two remaining normal
stresses $\sigma_{11}$ and $\sigma_{22}$, let us consider two stress tensors $\bsym{\Sigma}_{11} =
\diag(\sigma_{11},0,0)$ and $\bsym{\Sigma}_{22} = \diag(0,\sigma_{22},0)$ that impose these
loadings. We have shown above that the independence of the CRSS on the hydrostatic stress $\sigma_h$
and on the stress component $\sigma_{33}$ means that the CRSS can depend only on a stress tensor
that applies an equibiaxial tension-compression in the $xy$ plane. Hence, the relevant part of the
stress tensor $\bsym{\Sigma}_{11}$ is the deviatoric stress $\bsym{\Sigma}'_{11} =
\diag(\sigma_{11}/2,-\sigma_{11}/2,0)$. Similarly, only the deviatoric part $\bsym{\Sigma}'_{22} =
\diag(-\sigma_{22}/2,\sigma_{22}/2,0)$ of the stress tensor $\bsym{\Sigma}_{22}$ can affect the
dislocation glide. Comparisons of these stress tensors with $\bsym{\Sigma}_\tau$ above lead to the
following observations: (i) loading by the normal stress $\sigma_{11}$ is equivalent to imposing the
shear stress perpendicular to the slip direction $\tau=-\sigma_{11}/2$, and (ii) loading by the
normal stress $\sigma_{22}$ is equivalent to applying $\tau=\sigma_{22}/2$.

In order to substantiate these conclusions, we have calculated the variation of the CRSS with the
stress components $\sigma_{11}$ and $\sigma_{22}$ and compared them to the dependencies of the CRSS
on $\tau$ obtained earlier \cite{groger:08a}. These calculations were done similarly as above, when
investigating the effects of $\sigma_h$ and $\sigma_{33}$. The obtained dependencies of CRSS on
$\sigma_{11}$ and $\sigma_{22}$ are plotted in \reffig{fig:CRSS-all} by the red and green empty
triangles, respectively. The CRSS values obtained by applying the stress tensors
$\bsym{\Sigma}'_{11}$ and $\bsym{\Sigma}'_{22}$, plotted in \reffig{fig:CRSS-all} by the red and
green filled triangles, respectively, are derived from the CRSS vs. $\tau$ data that are plotted in
this figure by the black squares. One can clearly see that the dependencies in red and green plotted
in \reffig{fig:CRSS-all} by like colors are essentially identical\footnote{The small differences in
  the CRSS values are comparable with the error bars of the incremental estimate of the CRSS by
  atomistic simulations.}, which validates our observations above.

%Moreover, the dislocation moves on the $(\bar{1}01)$ plane when applying the stress tensor
%$\bsym{\Sigma}_{11}$ or $\bsym{\Sigma}_{22}$ with $|\sigma_{11}|/C_{44}\leq 0.03$ or
%$|\sigma_{22}|/C_{44}\leq 0.03$, respectively. The same slip plane is observed when applying the
%stress tensor $\bsym{\Sigma}_\tau$ with $|\tau|/C_{44}\leq 0.015$.

\begin{figure}[!htb]
  \centering
  \includegraphics[scale=0.19]{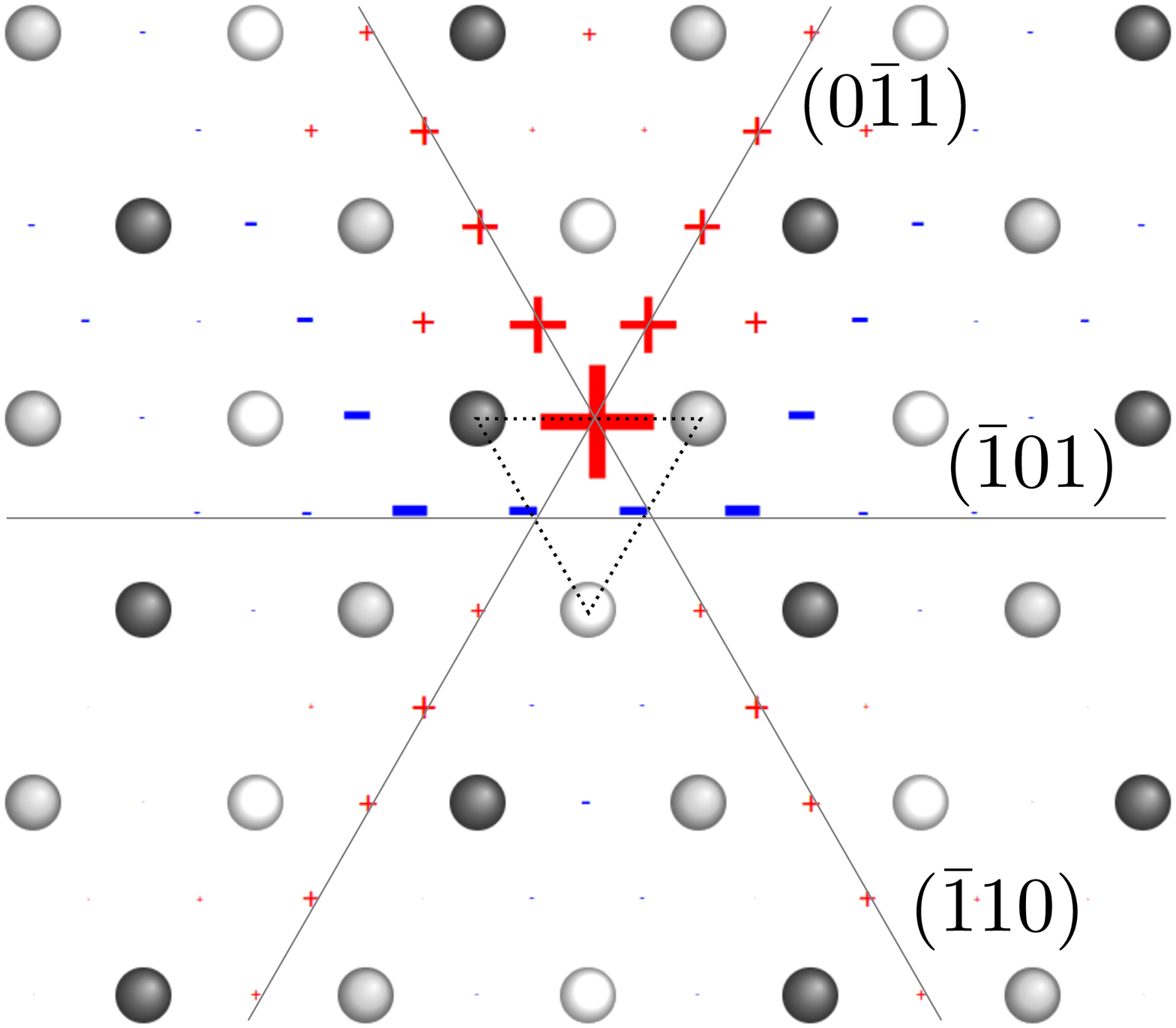} \quad
  \includegraphics[scale=0.19]{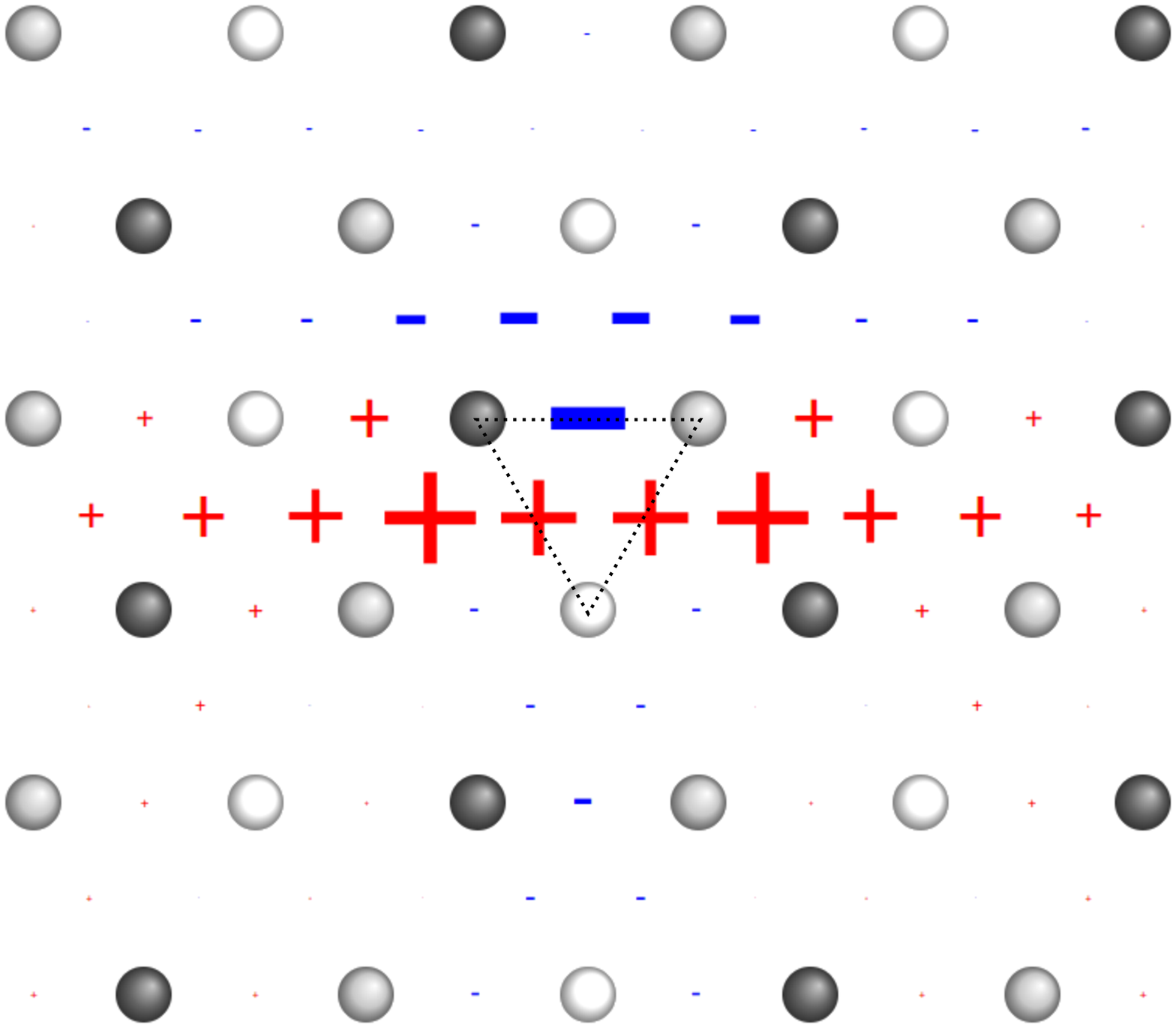} \\
  {\footnotesize (a) \hskip4cm (b)} \\
  \caption{Changes of differential displacement maps upon applying the stresses (a)
    $\sigma_{11}=0.03C_{44}$, and (b) $\sigma_{22}=0.03C_{44}$. The plane of the figure coincides
    with the $(111)$ plane and the lattice site that contains the center of the dislocation is
    marked by the triangle. The red ``$+$'' (blue ``$-$'') symbols represent the increase (decrease)
    of the screw components of the differential displacement map upon incorporating the stress into
    the relaxed block with the dislocation. If $\sigma_{22}<0$, the core changes are the same as
    those in (a), while for $\sigma_{11}<0$ they are indistinguishable from the figure (b).}
  \label{fig:dcore}
\end{figure}

For completeness, the plots in \reffig{fig:dcore} show the regions in which the screw component of
the differential displacement map \cite{vitek:70} is increased (red ``+'' symbols) and decreased
(blue ``--'' symbols) upon applying positive stresses $\sigma_{11}$ and $\sigma_{22}$ using the
stress tensors $\bsym{\Sigma}_{11}$ and $\bsym{\Sigma}_{22}$, respectively. The size of each symbol
represents the magnitude of an increase or a decrease of this screw component. It follows from
\reffig{fig:dcore}a, that the effect of $\sigma_{11}>0$ is to extend the dislocation core on the
$(0\bar{1}1)$ and $(\bar{1}10)$ planes and constrict it on the $(\bar{1}01)$ plane. We have
demonstrated already in Ref.~\cite{groger:08a} that the same distortion of the dislocation core is
caused by a negative applied shear stress $\tau$. Similarly, \reffig{fig:dcore}b shows that the
effect of $\sigma_{22}>0$ is to extend the dislocation core on the $(\bar{1}01)$ plane and constrict
it on the other two $\gplane{110}$ planes of the $[111]$ zone. Again, this agrees with the effect of
positive $\tau$ on the dislocation core, as discussed in Ref.~\cite{groger:08a}. The comparisons
above provide ample evidence that the normal stresses $\sigma_{11}$ and $\sigma_{22}$ should not be
thought of as acting in isolation. Instead, they act concomitantly by imposing a equibiaxial
tension-compression in the $xy$ plane and manifest themselves by the dependence of the CRSS on
the stress $\tau$ that induces shear in the plane perpendicular to the slip direction.

\subsection{Dependence of the CRSS on stress combinations}

In the following, we will investigate a combined effect of $\tau$ and $\sigma_{33}$ on the CRSS.
The initial simulation blocks for these calculations have been obtained by first applying the stress
$\tau$ in steps, as before. By keeping this stress constant, we then imposed the stress
$\sigma_{33}$ in the same steps until the block was subjected to the desired combination
$\{\tau,\sigma_{33}\}$. Starting with this stressed block, the shear stress parallel to the slip
direction ($\sigma_{23}$) was applied as before until the dislocation moved; this value of
$\sigma_{23}$ then represents the CRSS for a given combination of $\tau$ and $\sigma_{33}$. These
calculations have been made for the values of $\tau/C_{44} = \{0, \pm0.01, \pm0.02, \pm0.03\}$ and
$\sigma_{33}/C_{44} = \{0, \pm0.03\}$. It is important to emphasize that each stress state that
imposes a nonzero value of $\sigma_{33}$ applies at the same time a hydrostatic stress
$\sigma_{h}=\sigma_{33}/3$.

\begin{figure}[!htb]
  \centering
  \includegraphics[scale=0.33]{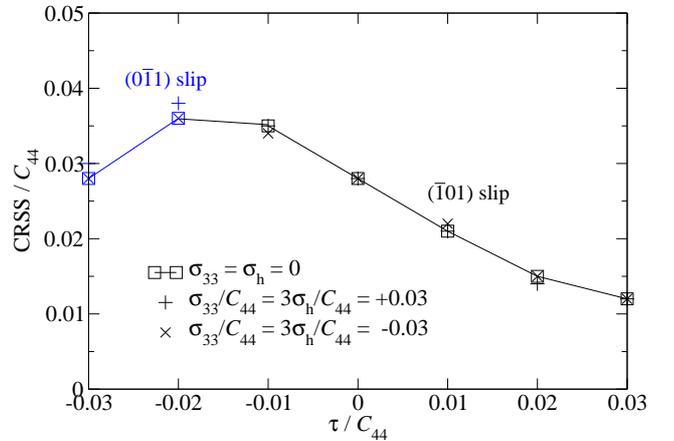}
  \caption{Dependence of the CRSS on the shear stress perpendicular to the slip direction ($\tau$)
    for three values of the stress component $\sigma_{33}$ (and the corresponding hydrostatic stress
    $\sigma_h$). The black symbols represent a dislocation glide on the $(\bar{1}01)$ plane, while
    the blue symbols a glide on the $(0\bar{1}1)$ plane.}
  \label{fig:CRSS-tau-sig33}
\end{figure}

The obtained dependencies of the CRSS on $\tau$ for the three values of $\sigma_{33}$ (and the
corresponding $\sigma_h$) are plotted in \reffig{fig:CRSS-tau-sig33}. This figure clearly shows that
the presence of the stress $\sigma_{33}$ has no effect on the CRSS vs. $\tau$ dependence. This
observation further proves our assertion that the stress components $\sigma_{33}$ and $\sigma_h$ do not
play any role in the glide of $1/2\gdir{111}$ screw dislocations in bcc tungsten. This conclusion
should be valid generally for all bcc metals and most likely also for other materials in which the plastic
deformation is governed by the dislocations that possess non-planar cores.

\section{Orientational effects of the CRSS and $\tau$}

In the calculations above, we considered that the MRSSP coincides with the $(\bar{1}01)$ plane,
which is the $\gplane{110}$ plane of the $[111]$ zone with the highest Schmid stress. Similar
calculations can be carried out for any orientation of the MRSSP, i.e. for any angle $\chi$ that the
MRSSP makes with the $(\bar{1}01)$ plane. The loading is then defined in the right-handed orthogonal
coordinate system in which the $y'$ axis is perpendicular to the MRSSP and the $z'$ axis parallel to
the $[111]$ direction. Any applied load can be expressed in this system as a combination of the
stress $\tau$ that imposes shear in the plane perpendicular to the slip direction and
the shear stress $\sigma$ parallel to the slip direction,
\begin{equation}
  \bsym{\Sigma}^{\MRSSP} = \left[
    \begin{array}{ccc}
      -\tau & 0 & 0 \\
      0 & \tau & \sigma \\
      0 & \sigma & 0
    \end{array} 
    \right] \ .
  \label{eq:SigmaMRSSP}
\end{equation}
In atomistic simulations, the stress tensor above has to be resolved in the orientation of the
atomic block, where the $y$ axis is perpendicular to the $(\bar{1}01)$ plane and $z \equiv z'$. This
transformation constitutes a simple rotation of the coordinate system by $-\chi$ around the $[111]$
axis. The transformed stress tensor
\begin{equation}
  \bsym{\Sigma}^{(\bar{1}01)} = \left[
    \begin{array}{ccc}
      -\tau\cos 2\chi & \tau\sin 2\chi & \sigma\sin\chi \\
       \tau\sin 2\chi & \tau\cos 2\chi & \sigma\cos\chi \\
       \sigma\sin\chi & \sigma\cos\chi & 0
    \end{array} 
    \right]
  \label{eq:Sigma110}
\end{equation}
is then used to impose the applied load by displacing all atoms in the simulated cell. In the
following, the superscript will refer to the coordinate system in which the stress component is
resolved. For example, $\sigma_{ij}^{(hkl)}$ refers to a right-handed orthogonal coordinate system
with the $z'$ axis parallel to the $[111]$ direction, and the $y'$ axis parallel to the $[hkl]$
direction.

One can immediately see that for $\chi\not=0$, the shear stress $\sigma$ parallel to the slip
direction and applied in the MRSSP acts in the orientation of the simulated block by the glide
stress $\sigma_{23}^{(\bar{1}01)}=\sigma\cos\chi$ and by the non-glide stress
$\sigma_{13}^{(\bar{1}01)}=\sigma\sin\chi$. If the dislocation glide was not affected by the stress
component $\sigma_{13}^{(\bar{1}01)}$, the CRSS would be proportional to $1/\cos\chi$ and would thus
be symmetric about $\chi=0$. This is not the case in bcc metals \cite{ito:01,groger:08a,groger:09b}
and the observed twinning-antitwinning asymmetry of the CRSS is attributed to the effect of the
stress component $\sigma_{13}^{(\bar{1}01)}$. This implies that two terms are needed in the yield
criterion to describe the orientational dependence of the shear stress parallel to the slip
direction.

\begin{figure}[!htb]
  \centering
  \includegraphics[scale=0.41]{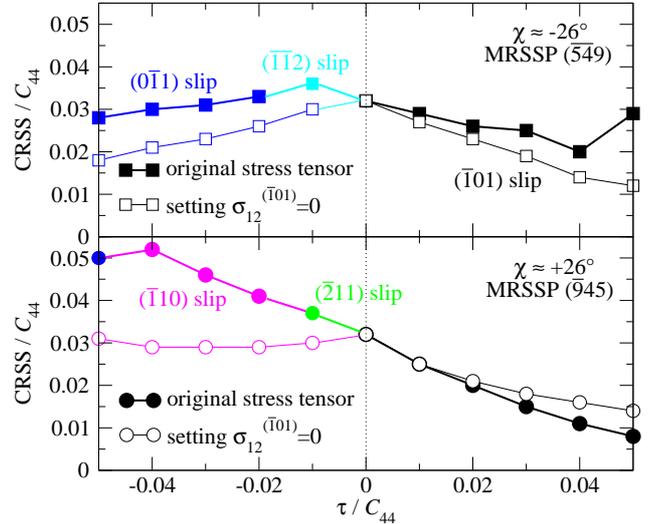}
  \caption{The CRSS vs. $\tau$ dependencies calculated for the MRSSPs with angles $\chi \approx
    -26^\circ$ (upper panel) and $\chi \approx +26^\circ$ (lower panel). The filled symbols (thick
    lines) correspond to the data obtained using the full stress tensor \refeq{eq:Sigma110}, while
    the empty symbols (thin lines) are obtained using \refeq{eq:Sigma110} in which
    $\sigma_{12}^{(\bar{1}01)}$ is artificially set to zero. The colors distinguish different slip
    planes in the $[111]$ zone; the two $\gplane{112}$ slips are composed of alternating steps of
    the dislocation on two adjacent $\gplane{110}$ planes.}
  \label{fig:chi26}
\end{figure}

Similarly, for $\chi\not=0$, the stress $\tau$ acts by a pair of normal stresses
$\sigma_{22}^{(\bar{1}01)}=-\sigma_{11}^{(\bar{1}01)}=\tau\cos 2\chi$ and by the shear stress
$\sigma_{12}^{(\bar{1}01)}=\tau\sin 2\chi$. The effect of the latter is to modify the $\CRSS$
vs. $\tau$ dependence as the angle of the MRSSP deviates from $\chi=0$ and, possibly, to alter the
slip plane on which the dislocation moves. To demonstrate the effect of the shear stress
$\sigma_{12}^{(\bar{1}01)}$, we will consider in the following the MRSSPs $(\bar{9}45)$ and
$(\bar{5}\bar{4}9)$ that make angles $\chi \approx \pm 26^\circ$ with the $(\bar{1}01)$
plane. Instead of calculating the CRSS vs. $\tau$ dependence using the full stress tensor
\refeq{eq:Sigma110}, we will now artificially set the component $\sigma_{12}^{(\bar{1}01)}$ to zero.
The obtained data are plotted in \reffig{fig:chi26} by empty symbols and interpolated by thin
lines. For comparison, our previous results \cite{groger:08a} obtained using the full stress tensor
\refeq{eq:Sigma110} are plotted in this figure by filled symbols (thick lines). The observed
differencies of the CRSS at both positive and negative $\tau$ reveal that the stress component
$\sigma_{12}^{(\bar{1}01)}$ cannot be neglected. Its presence promotes the composite (or zig-zag)
slip of the dislocation on two $\gplane{110}$ planes around $\tau/C_{44}=-0.01$, which gives rise to
an average slip plane of the $\gplane{112}$ type. These observations suggest that the CRSS and the
orientation of the actual slip plane depend both on the magnitude of $\tau$ and on the orientation
of the MRSSP.

The analysis above implies that the yield criterion for bcc metals has to contain four stress
components. Two of these are the shear stresses parallel to the slip direction resolved in two
arbitrary (but non-coplanar) planes of the $[111]$ zone. In the yield criterion that we have
developed in Ref.~\cite{groger:08b}, the shear stress parallel to the slip direction was resolved in
the $(\bar{1}01)$ and $(0\bar{1}1)$ planes as $\sigma_{23}^{(\bar{1}01)} = \CRSS \cos\chi$ and
$\sigma_{23}^{(0\bar{1}1)} = \CRSS \cos(\chi+\pi/3)$. The effect of the shear stress perpendicular
to the slip direction, investigated in the atomistic simulations by applying the equibiaxial
tension-compression, i.e. $\tau$ in Eq.~\refeq{eq:SigmaMRSSP}, is incorporated in the yield
criterion \cite{groger:08b} by two shear stresses $\sigma_{12}$ acting in the same two
$\gplane{110}$ planes as above. In particular, $\sigma_{12}^{(\bar{1}01)} = \tau \sin 2\chi$ and
$\sigma_{12}^{(0\bar{1}1)} = \tau \cos(2\chi+\pi/6)$. Nevertheless, it should be emphasized that
these planes need not be the same as those considered above when describing the orientational
dependence of the shear stress parallel to the slip direction. The simplest yield criterion for bcc
metals can thus be written as a linear combination of the four stress components identified above,
i.e.
\begin{equation}
  \sigma_{23}^{(\bar{1}01)} + a_1\sigma_{23}^{(0\bar{1}1)} + 
  a_2\sigma_{12}^{(\bar{1}01)} + a_3\sigma_{12}^{(0\bar{1}1)} = \tau^*_{cr}  \ . \ \ \
\end{equation}
The coefficients $a_1$, $a_2$, $a_3$ and the critical stress $\tau^*_{cr}$ are obtained in
Ref.~\cite{groger:08b} by fitting the atomistically calculated data of CRSS vs. $\chi$ and CRSS
vs. $\tau$ for a number of orientations of the MRSSP.

\section{Conclusions}

The objective of this paper has been to identify the stress components that affect the CRSS to move
an isolated $1/2[111]$ screw dislocation in bcc metals in molecular statics calculations. We have
shown that the CRSS does not depend on the hydrostatic stress and the stress component $\sigma_{33}$
that is parallel to the dislocation line. The latter observation contradicts the conclusions made in
Ref.~\cite{koester:12}. It thus follows that the Peierls stress cannot depend directly on the
two remaining normal stresses $\sigma_{11}$ and $\sigma_{22}$ but only on their combination
represented by the stress tensor $\bsym{\Sigma}_\tau=\diag(-\tau,\tau,0)$, which applies the shear
stress perpendicular to the slip direction. The importance of this stress component has been
recognized already by Ito and Vitek \cite{ito:01} and quantified fully for bcc molybdenum and
tungsten in Ref.~\cite{groger:08a}. Here, we were able to reproduce the trend in the dependence of
the CRSS on the stress components $\sigma_{11}$ and $\sigma_{22}$, calculated for bcc iron by
Koester et al. \cite{koester:12}. However, owing to the observation above, we conclude that these
dependencies are the manifestations of the dependence of the CRSS on the shear stress perpendicular
to the slip direction.

The conclusions of the work of Koester et al. \cite{koester:12} have been recently adopted by Lim
et al. \cite{lim:13} to develop a yield criterion and a crystal plasticity finite element model for
various bcc metals both of which explicitly contain the effects of normal stresses $\sigma_{11}$,
$\sigma_{22}$, and $\sigma_{33}$. They argue that the yield criterion that we have developed in
Ref.~\cite{groger:08b} to describe the onset of yielding in bcc molybdenum and tungsten and which
follows from the non-associated flow theory developed in Refs.~\cite{qin:92, qin:92b} has to be
augmented by other terms that incorporate the effects of the three normal stresses, while keeping
the hydrostatic stress zero. However, the calculations made in this paper provide a convincing proof
that there is no influence of $\sigma_{33}$ on the CRSS, while both remaining normal stresses act
via the shear stress perpendicular to the slip direction.

We have shown in this paper that only four stress components affect the glide of $1/2\gdir{111}$
screw dislocations in bcc metals for any orientation of the MRSSP. Two of these are the shear
stresses $\sigma_{23}$ parallel to the slip direction resolved in two different planes of the
$\gdir{111}$ zone. Their role is to represent the twinning-antitwinning asymmetry of the CRSS. The
other two are the shear stresses $\sigma_{12}$ perpendicular to the slip direction, again acting in
two different planes of the $\gdir{111}$ zone but not necessarily the same as the two shear stresses
parallel to the slip direction above. The shear stresses perpendicular to the slip direction affect
not only the CRSS but also the slip plane on which the dislocation moves at larger negative values
of $\tau$. The onset of yielding in bcc single crystals can thus be described by yield criteria that
involve linear combinations of these four stress components, as demonstrated for bcc molybdenum and
tungsten in Ref.~\cite{groger:08b}. The calculations made in this paper prove that these yield
criteria are complete and, contrary to the assertions of Koester et al. \cite{koester:12} and Lim et
al. \cite{lim:13}, there are no effects of other stress components besides those considered in the
formulation developed in Ref.~\cite{groger:08b}.

\section{Acknowledgments}

Fruitful discussions on the topic with Vaclav Vitek and his support and encouragement over the past
decade are greatly appreciated. Support from the Marie-Curie International Reintegration Grant
No. 247705 ``MesoPhysDef'', and a partial support from the Academy of Sciences of the Czech
Republic, Project no. RVO:68081723 are acknowledged. The access to the MetaCentrum computing
facilities provided under the program ``Projects of Large Infrastructure for Research, Development,
and Innovations'' LM2010005 funded by the Ministry of Education, Youth, and Sports of the Czech
Republic is appreciated. This work has been carried out at the Central European Institute of
Technology (CEITEC) with research infrastructure supported by the project CZ.1.05/1.1.00/02.0068
financed from the EU Structural Funds.

\bibliographystyle{elsarticle-num}
\bibliography{bibliography}

\end{document}